\documentclass[fleqn,twoside]{article}
\usepackage{espcrc2}

\usepackage{amssymb}

\usepackage{graphicx}
\usepackage[figuresright]{rotating}

\newcommand{\AmS}{{\protect\the\textfont2
  A\kern-.1667em\lower.5ex\hbox{M}\kern-.125emS}}

\title{\hfill\begin{minipage}{0pt}\scriptsize \begin{tabbing}
       \hspace*{\fill} Edinburgh-2004/14\\
\end{tabbing}
\end{minipage}\\[8pt]
       \vspace{-0.3cm}
Exploratory spectrum calculations using overlap valence quarks on a staggered sea}

\author{UKQCD Collaboration, K.C. Bowler\address[MCSD]{School of Physics, The University of Edinburgh, 
        Edinburgh EH9 3JZ, UK}, 
	B. Jo\'o\addressmark,     
	R.D. Kenway\addressmark,
        C.M. Maynard\addressmark \ 
        and
        R.J. Tweedie\thanks{Talk presented by R.J. Tweedie}\addressmark[MCSD]}
       
\begin{document}

\begin{abstract}
We present exploratory results for the hadron mass spectrum and pseudoscalar meson decay constants using mixed actions. We use improved staggered sea quarks and HYP-smeared overlap valence quarks. We obtain good signals on 10 configurations at one lattice spacing and two different sets of sea quark masses.
\vspace{0.6pc}
\end{abstract}

\maketitle

\section{INTRODUCTION}

Solving lattice QCD to high precision requires the use of dynamical Ginsparg-Wilson light quarks. However, this is computationally expensive. Therefore, as a starting place, we take improved staggered quark configurations, which have the advantage of light sea quarks, and use an overlap valence quark action, which has the correct chiral and flavour symmetries. This has its own disadvantages. Firstly, the overlap inversion is still relatively expensive and, secondly, because we have different actions for the sea and valence quarks, it is not straightforward to interpret the results.

Mixed actions are inevitable in the improved staggered programme, because we do not yet have a local version of the sea quark action to use for the valence quarks. Also, it is complicated to measure some quantities in the improved staggered formalism. For example, in order to measure the mass of the nucleon one has a choice of O(100) different nucleon operators. Without measuring a representative set, the effect of taste symmetry breaking is unquantified and uncontrolled.


\section{SIMULATION PARAMETERS}

The simulations were performed on twenty coarse (MILC) dynamical configurations with 2+1 flavours \cite{Bernard:2001av}. Ten configurations have a light isodoublet with mass $m_{l}=\frac{3}{5}m_{s}$ and ten have $m_{l}=\frac{2}{5}m_{s}$. Both have a lattice spacing $a \simeq 0.125$ fm and linear size $L \simeq 2.5$ fm. Three iterations of HYP-smearing were applied to each configuration \cite{Hasenfratz:2001hp}. The overlap operator from SZIN code \cite{SZIN} was then used to calculate propagators. These were created with seven different valence quark masses using the overlap multi-mass solver: four light and three heavy \cite{Dong:2001fm}. Even for the baryon spectrum we get a remarkably good signal on ten configurations. 
\vspace{-0.1cm}
\subsection{HYP-smearing}

As a check, we applied several iterations of HYP-smearing to 624 quenched UKQCD configurations at $\beta=5.93$ with a volume of $16^{3}\times32$. Planar Wilson loops were used to extract the quark-antiquark potential.
HYP-smearing quickly alters the short-distance behaviour, while the medium-to-long distance behaviour remains relatively unchanged for a small number ($\lesssim 3$) of iterations. Smearing the configuration helps to speed up the convergence of the overlap operator, renders the effective interaction more local, and makes the eigenvalue spectrum more like that for an overlap sea \cite{Durr:2004as}.

\section{RESULTS}

\begin{figure}[htb]
\includegraphics[scale=0.3]{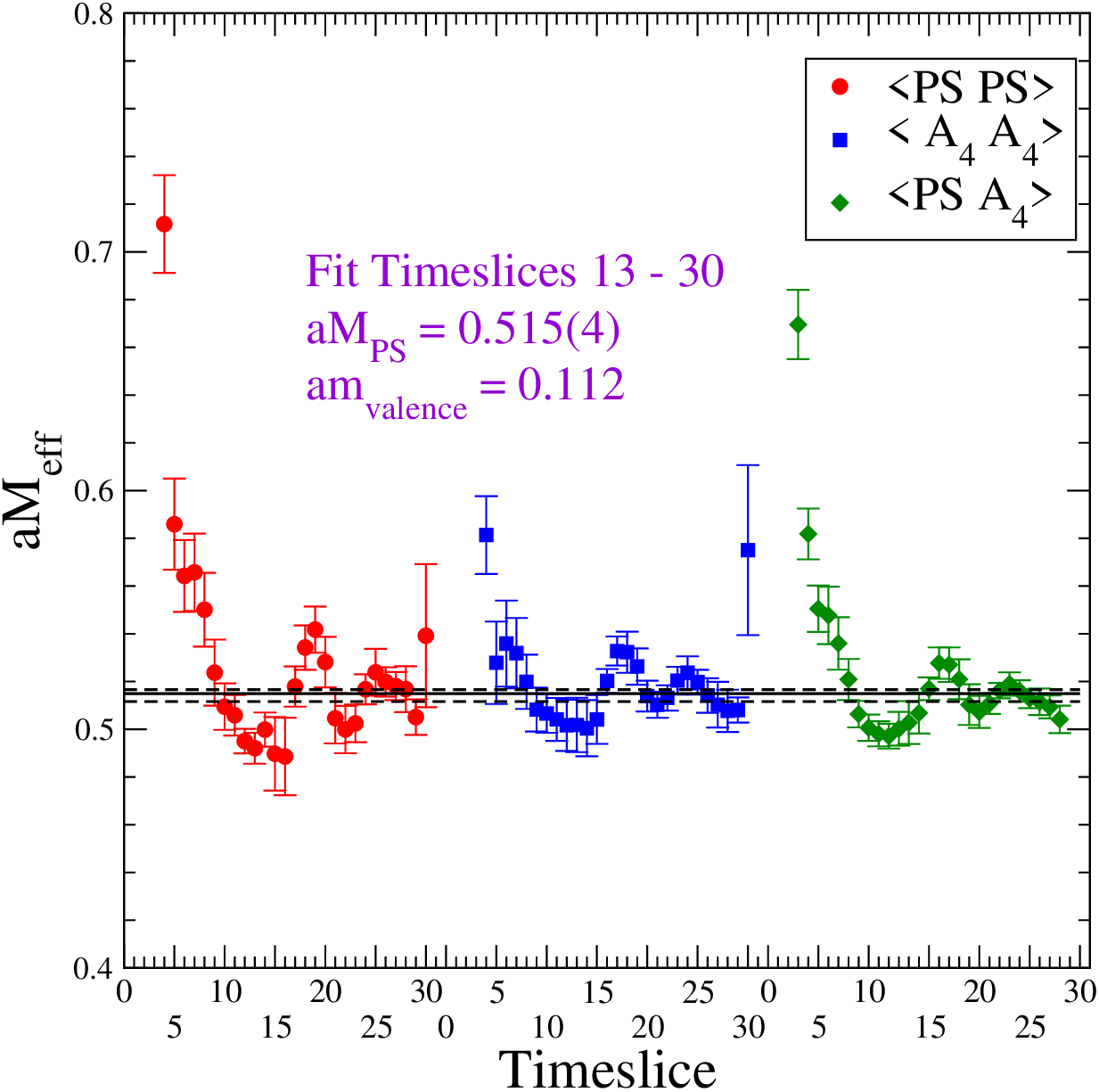}
\caption{Pseudoscalar meson effective mass and simultaneous uncorrelated fit to three correlators.}
\label{effective_mass}
\end{figure}

We perform simultaneous fits to three different correlators in order to extract the pseudoscalar meson mass (see figure \ref{effective_mass}).
The fluctuations in a$M_{\rm{eff}}$ are larger than the apparent statistical errors, but this is probably due to underestimation of the variance on ten configurations.

All the analysis is carried out partially quenched - we hold the sea quark mass fixed and vary the input valence quark mass. Since we have multiple input valence masses, we can construct non-degenerate light meson correlators. A two-dimensional fit was performed to $(aM_{PS})^{2}$ versus valence masses $m_{q_{1}}$ and $m_{q_{2}}$, which allowed evaluation of the average $u$ and $d$ quark mass, $\hat{m}$, from
\begin{equation}
M_{\pi}^{2} = B\left(m_{q_{1}}+m_{q_{2}}\right) + A = 2B\hat{m} + A 
\end{equation}
where $M_{\pi}$ is the physical pion mass squared.
 
This in turn allows us to evaluate the strange quark mass from
\begin{equation}
M_{K}^{2} = B\left(m_{s}+\hat{m}\right) + A 
\end{equation}
where $M_{K}$ is the physical kaon mass and $m_{s}$ is the
strange quark mass.

\subsection{Pseudoscalar Meson Decay Constants}
\begin{figure}[htb]
\includegraphics[scale=0.3]{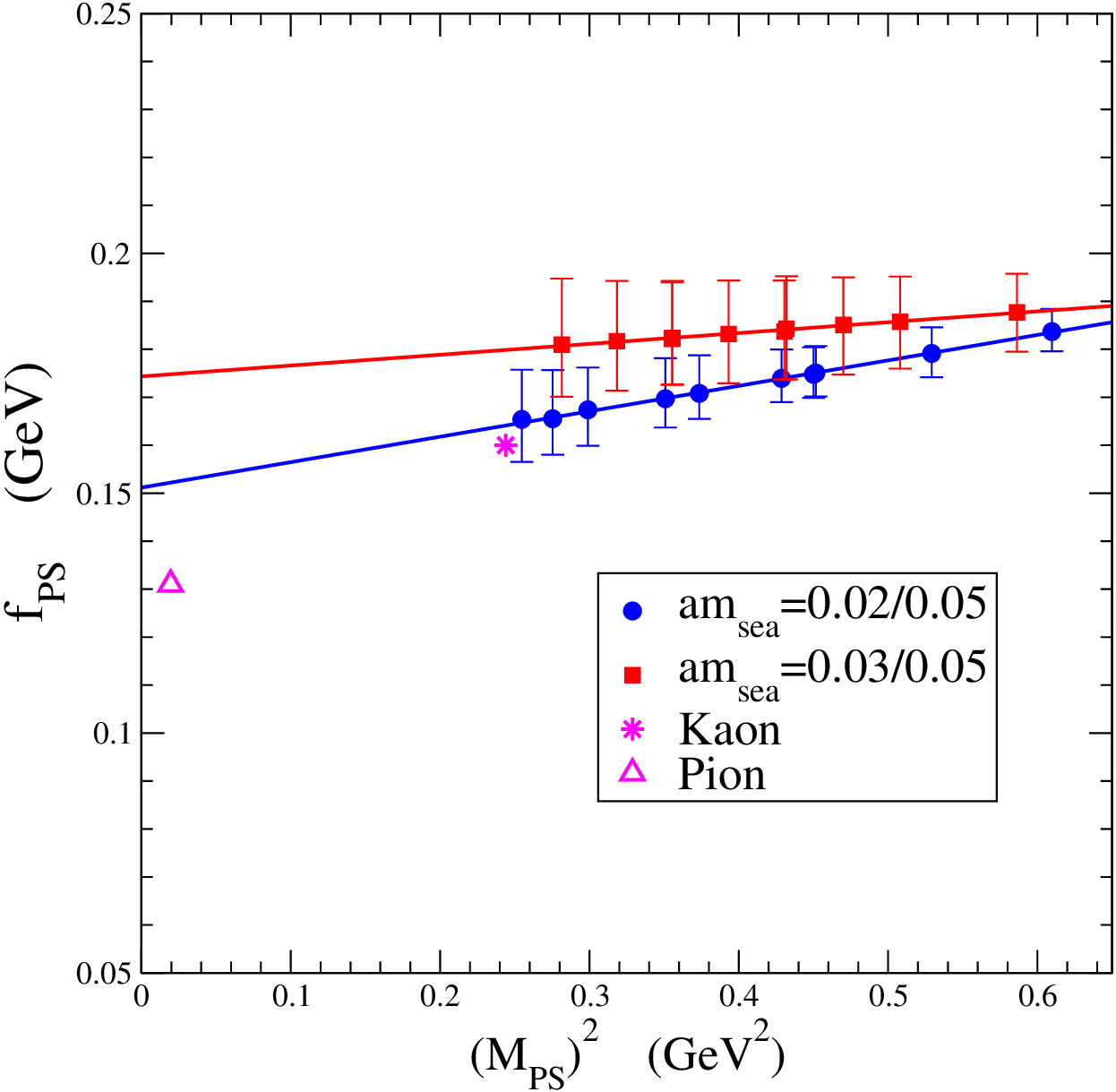}
\vspace{-0.5cm}
\caption{$f_{PS}$ versus $M_{PS}^{2}$ for the two ensembles.}
\label{Pseudoscalardecay}
\end{figure}

We define $f_{PS}$ as
\begin{equation}\label{fPS}
f_{PS} = \frac{Z_{A}\langle0|A_{4}|PS\rangle}{M_{PS}} .
\end{equation}
We obtain $Z_{A}$ from the axial Ward identity
\begin{equation}
Z_{A}\langle\partial_{\mu}A_{\mu} \mathcal{O} \rangle = 2m_{q}\langle P \mathcal{O} \rangle 
\end{equation}
which we can express in terms of the pseudoscalar correlator, $C_{PP}$, and the
pseudoscalar axial correlator, $C_{PA_{4}}$. $\langle0|A|PS\rangle$ cancels in eq.(\ref{fPS}) and hence we only require $C_{PP}$ in order to evaluate
$f_{PS}$. Once again we perform a 2-d linear fit to the light
non-degenerate pseudoscalars to calculate $f_{PS}$ (see figure
\ref{Pseudoscalardecay}) and extract the ratio of
$\frac{f_{K}}{f_{\pi}}$ (see table \ref{table1}). The value
increases slightly with decreasing light sea quark mass in the right direction to agree with experiment. This is also evident from the slight change of the gradient in figure \ref{Pseudoscalardecay}.
\vspace{-0.2cm}

\begin{table}[htb]
\caption{Pseudoscalar Meson Decay constants}
\label{table1}
\newcommand{\m}{\hphantom{$-$}}
\newcommand{\cc}[1]{\multicolumn{1}{c}{#1}}
\renewcommand{\tabcolsep}{1pc} 
\renewcommand{\arraystretch}{1.2} 
\begin{tabular}{@{}lll}
\hline
Sea Quarks            & $f_{K}/f_{\pi}$ & $f_{Ds}$   (MeV)  \\
\hline
$am_{\rm{sea}}$ = 0.03/0.05     	  & \m1.03(3) & \m226(14)\\
$am_{\rm{sea}}$ = 0.02/0.05                & \m1.08(4) & \m232(11)\\
\hline
Expt: \cite{Eidelman:2004wy} 		    & \m1.22(1) & \m 266(32)\\
\hline
\end{tabular}\\[2pt]
\end{table}
\vspace{-0.5cm}
\subsection{Baryon Spectrum}

\begin{figure}[htb]
\includegraphics[scale=0.3]{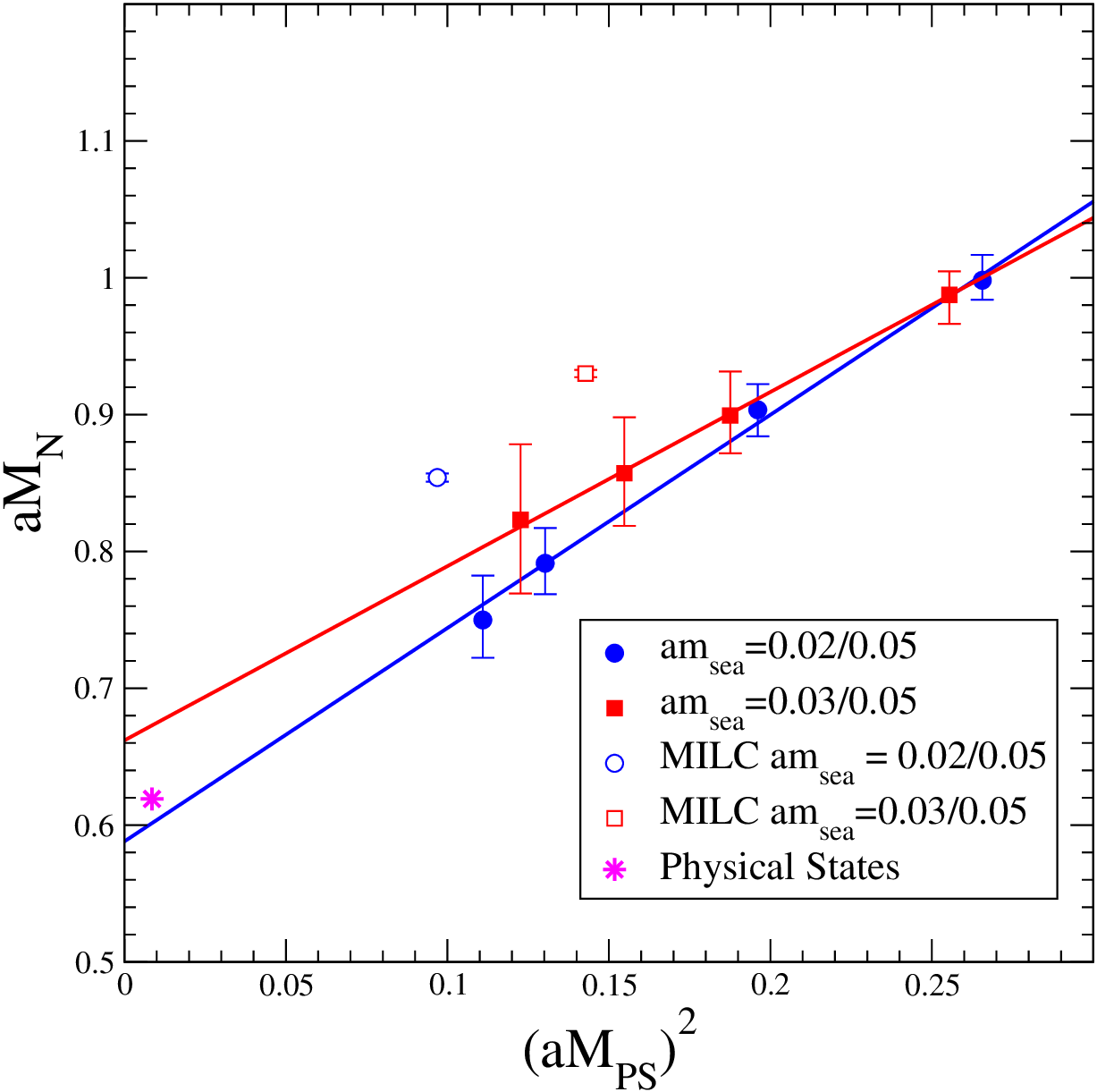}
\vspace{-0.2cm}
\caption{Nucleon mass vs pseudoscalar meson mass squared.}
\label{nucleonlinear}
\end{figure}

We measure the masses of the nucleon and delta baryon. It is remarkable that we can see a signal for the negative parity partner of the nucleon on as few as ten configurations. Figure \ref{nucleonlinear} shows the nucleon mass versus the pseudoscalar meson mass squared. The lines shown are uncorrelated linear fits to our data. The values calculated by the MILC collaboration \cite{Bernard:2001av} on their corresponding full ensembles are also shown. 
\vspace{-0.5cm}
\subsection{Charm Physics}

Heavy quarks essentially come for free in the overlap propagator calculation due to the multi-mass solver. However, lattice artefacts are $\cal{O}$$(am_{q})^{2}$ and the heaviest input valence quark mass used is $am_{q} = 0.84$ and hence $(am_{q})^{2} \sim 0.7$. With the lattice spacing of $a^{-1} \sim 1.5$GeV, we are at best on the limit of simulating charm. Because of the rapid decay in Euclidean time, we require double precision. However, this does not slow the solver down appreciably as we need less reorthogonalisations of the Krylov subspace than in single precision.

\begin{figure}[htb]
\includegraphics[scale=0.3]{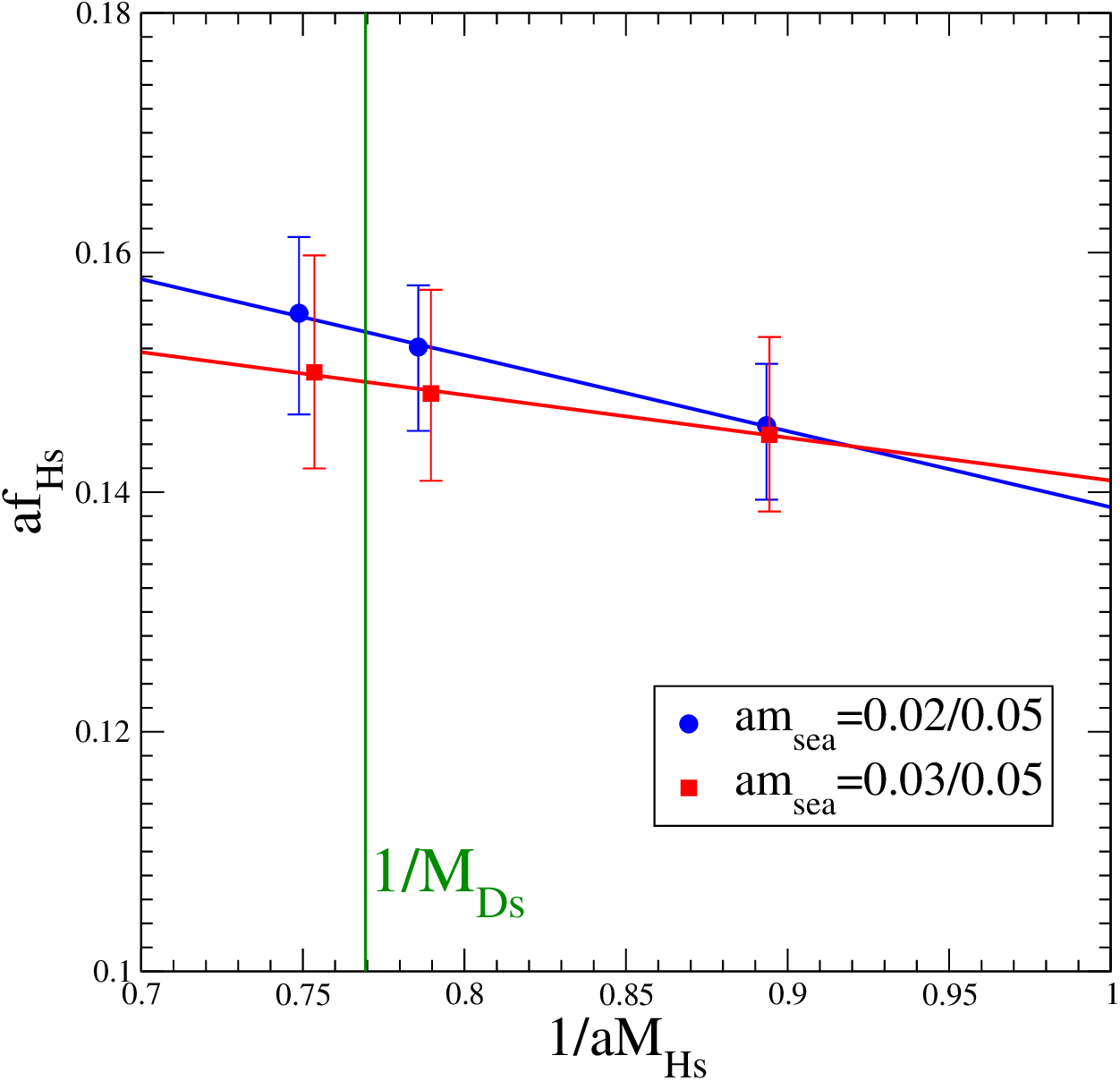}
\vspace{-0.3cm}
\caption{$f_{Hs}$ vs inverse heavy-strange pseudoscalar meson mass.}
\label{fDs}
\end{figure}

These heavy quark propagators were used to calculate $f_{Ds}$ (see table \ref{table1}). 
The value of $f_{Ds}$ increases with decreasing light sea quark mass, in the direction of the experimental value, as can be seen from the change of gradients in figure \ref{fDs}. 

\vspace{-1.2cm}
\section{SUMMARY}

We have shown that, even with very low statistics, it is possible to calculate light hadron and charm physics using an overlap valence operator on an improved staggered sea. This could be an alternative to staggered valence quarks for probing sea quark effects, although for our data, the effect of small changes in the sea quark mass is minimal. 


\normalsize
\end{document}